\documentclass[a4paper,11pt]{article}
\usepackage{jcappub,subfigure} 
\usepackage{lineno}
\usepackage{rotating}
\usepackage{orcidlink}

\arxivnumber{2508.19081} 
\title{\boldmath Lensing amplitude anomaly and varying electron mass alleviate the Hubble and $S_8$ tensions}

\author[a]{Yi-Ying~Wang\,\orcidlink{0000-0003-1215-6443},}
\author[a, b]{Lei~Lei\,\orcidlink{0000-0003-4631-1915},}
\author[a]{Shao-Peng~Tang\,\orcidlink{0000-0001-9120-7733},}
\author[a, b]{Yi-Zhong~Fan\,\orcidlink{0000-0002-8966-6911}\footnote{Corresponding author}}

\affiliation[a]{Key Laboratory of Dark Matter and Space Astronomy, Purple Mountain Observatory, Chinese Academy of Sciences, Nanjing 210033, People's Republic of China}
\affiliation[b]{School of Astronomy and Space Science, University of Science and Technology of China, Hefei, Anhui 230026, People's Republic of China}

\date{\today}
\emailAdd{wangyy@pmo.ac.cn}
\emailAdd{leilei@pmo.ac.cn}
\emailAdd{tangsp@pmo.ac.cn}
\emailAdd{yzfan@pmo.ac.cn}

\abstract{
Cosmological measurements have revealed tensions within the standard $\Lambda$CDM model, notably discrepancies in the Hubble constant and $S_8$ parameter. A modified recombination scenario involving a time-varying electron mass has been proposed as a feasible solution to the Hubble tension without exacerbating the $S_8$ tension. Recent observations have further revealed other potential deviations from the $\Lambda$CDM framework, such as non-flat spatial curvature and an anomalous CMB lensing amplitude. In this study, we explore whether introducing a variation in the electron mass $m_e$, allowing non-zero spatial curvature $\Omega_K$, and a free lensing amplitude $A_{\rm lens}$ can resolve these persistent tensions. Using the Planck Public Release (PR) 3 and ACT power spectra, Planck PR4 and ACT lensing maps, together with BAO measurements from DESI DR2, we obtain $H_0 = 69.61^{+0.60}_{-0.55} \rm \, km \, s^{-1} \, Mpc^{-1}$ and $S_8= 0.808\pm0.012$, with $\Delta m_e / m_e = 0.0109^{+0.0068}_{-0.0066}$ and $A_{\rm lens} = 1.030^{+0.039}_{-0.037}$, both exceeding the $\Lambda$CDM expectations. We find no indication of spatial curvature deviating from flatness, even when including the Cosmic Chronometers and SNe Ia samples. However, when adopting the latest Planck power spectra likelihoods, NPIPE and HiLLiPoP, we obtain lower electron masses with $\Delta m_e / m_e = -0.0063^{+0.0095}_{-0.0099}$ and $-0.0095^{+0.0078}_{-0.0079}$, relieving the $S_8$ tension only. The lensing amplitude remains anomalously high, with $A_{\rm lens} = 1.053^{+0.042}_{-0.040}$ and $1.075^{+0.044}_{-0.043}$. Our results point to a promising direction for cosmological models to reconcile the aforementioned discrepancies, although more precise data from future experiments will be necessary to clarify the aforementioned modifications.}

\begin{document}

\maketitle
\flushbottom

\section{Introduction}
\label{sec:intro}
The $\Lambda$ Cold Dark Matter ($\Lambda$CDM) model serves as the standard cosmological paradigm, providing an impressively successful description of a wide range of observations with only six fundamental parameters. Nevertheless, the precise determination of several of these parameters remains the subject of considerable debate. On the one hand, notable $tensions$ have emerged between early- and late-time cosmological measurements. For instance, the local determination of the Hubble constant $H_0$ via the distance-ladder method \citep{2021ApJ...908L...6R,2022ApJ...934L...7R} differs from the value inferred from cosmic microwave background (CMB) observations \citep{2020A&A...641A...6P} at the $5\sigma$ level. Direct measurements of the growth rate of cosmological perturbations $S_8$ tend to favor values lower by $2$–$3\sigma$ \citep{2021A&A...646A.140H, 2022PhRvD.105b3520A} compared to the Planck 2018 $\Lambda$CDM prediction \citep{2020A&A...641A...6P} (It should be noticed that latest cosmic shear constraints from KiDs-Legacy show only a $0.73\sigma$ tension compared to the Planck 2018 results \citep{2025arXiv250319441W}.). On the other hand, several cosmological parameters exhibit deviations from their standard-model expectations. Examples include an anomalously high value of a phenomenological rescaling ($A_{\rm lens} > 1$) of the CMB lensing amplitude, which effectively modifies the lensing power spectrum from $C^{\Psi}_{\ell}$ to $A_{\rm lens}C^{\Psi}_{\ell}$, and a preference for a closed Universe ($\Omega_K < 0$) in Planck temperature and polarization data \citep{2020A&A...641A...6P, 2021PhRvD.103d1301H}. Hints of a time-dependent dark energy equation of state ($w \neq -1$) have also been reported from DESI baryon acoustic oscillation (BAO) measurements \citep{2025JCAP...02..021A, 2025PhRvD.112h3515A}. The potentially profound implications of these results have promoted a wide range of investigations into extensions of the $\Lambda$CDM paradigm, including new physics scenarios beyond the Standard Model (see refs.~\cite{2021CQGra..38o3001D, 2022JHEAp..34...49A, 2022NewAR..9501659P, 2023Univ....9...94H, 2024hct..book.....D, 2024ARA&A..62..287V, 2025PDU....4901965D} for reviews).

Furthermore, the spatial curvature of the Universe remains a topic of active debate. Using the official baseline \texttt{Plik} and \texttt{CamSpec} likelihoods from Planck 2018, ref.~\cite{2020NatAs...4..196D} found a strong degeneracy between $A_{\rm lens}$ and $\Omega_K$. Their analysis favored a closed Universe ($\Omega_K<0$) at greater than $99\%$ confidence level, though at the cost of exacerbating the $H_0$ tension due to the positive correlation between $H_0$ and $\Omega_K$ \citep{2025arXiv250926263S}. When Planck CMB lensing and BAO data were jointly analyzed, however, the reconstructed curvature parameter was consistent with a flat Universe, in line with $\Lambda$CDM expectations \citep{2020MNRAS.496L..91E}. After introducing cosmic chronometer measurements, spatial flatness was still supported \citep{2021ApJ...908...84V}. More recently, several studies confirmed the flatness of the curvature by multiple datasets \citep{2025JCAP...08..014C, 2025PDU....4901995B, 2025JCAP...08..018D}. For example, ref.~\cite{2024A&A...682A..37T} analyzed the latest Planck PR4 data release and obtained a mildly elevated lensing amplitude $A_{\rm lens} = 1.039\pm0.052$, along with a nearly flat curvature $\Omega_K=-0.012 \pm 0.010$, within a single-parameter extension to $\Lambda$CDM. Using the latest BAO measurements from DESI DR2, a flat spatial curvature was derived with $\Omega_K = 0.025\pm0.041$ \citep{2025PhRvD.112h3515A}. Meanwhile, evidences for a closed Universe were also reported from a range of other observations, including the nine-year WMAP data \citep{2013ApJS..208...20B}, cosmic chronometers and type Ia supernovae (SNe Ia) \citep{2021MNRAS.506L...1D, 2023MNRAS.523.3406F}, and combined full-shape galaxy clustering plus BAO measurements \citep{PhysRevD.103.023507}. Furthermore, an open Universe was argued based on similar datasets \citep{2021MNRAS.501.5714W, 2025PhRvD.112f3514W}, and even contradictory results emerged when applying different reconstruction techniques \citep{2023PhRvD.108f3522Q}.

Recent BAO measurements from DESI have spurred renewed interest in the nature of dark energy, particularly in dynamical dark energy (DDE) scenarios parameterized by the Chevallier–Polarski–Linder (CPL) model \citep{2001IJMPD..10..213C, PhysRevLett.90.091301}. Such models have also been invoked to explain the unexpectedly large abundance of massive galaxies observed at very high redshift \citep{2022ApJ...938L...5M, 2023JCAP...10..072A, 2024ApJ...976..227M, 2024EPJP..139..711W}. Ref.~\cite{2025PDU....4801906G} examined the statistical significance of the $w_0 w_a $CDM model using combined data sets, and found that the inclusion of Pantheon+ SNe Ia and SDSS BAO data substantially weakens the preference for dynamical evolution. Similarly, ref.~\cite{2025PhRvD.112b3508G} argued that CMB experiments other than Planck generally reduce the evidence for DDE, since Planck uniquely provides high-precision temperature and E-mode polarization anisotropy measurements at large angular scales. Additionally, in a $w_0w_a$CDM framework with a phantom-to-quintessence transition ($w_0 > -1$ and $w_a < 0$), the inferred value of $H_0$ is typically lower than in $\Lambda$CDM, thereby exacerbating the Hubble tension between early- and late-time measurements \citep{2018PhRvD..98h3501V, 2022JCAP...04..004L, 2024arXiv241212905C,  2025SCPMA..6880410P}.

These considerations suggest that new physics beyond the standard $\Lambda$CDM model may be required. Nevertheless, it remains an open question whether such modifications should be introduced in the late Universe or in the early Universe \citep{2020PhRvD.102j3525K, 2021MNRAS.504.3956A, 2022PhRvD.106f3519C, 2022NatAs...6.1484P, 2023Univ....9..393V}. For example, a local void \cite{2020MNRAS.499.2845H, 2025MNRAS.540..545B} and can both alleviate the Hubble tension. To quantify the relative efficacy of different $H_0$ tensions, ref.~\cite{2022PhR...984....1S} systematically evaluated 16 extensions to $\Lambda$CDM, including scenarios with dark radiation, early-Universe solutions, and late-Universe solutions. They found that a model combining spatial curvature and a shift in the recombination epoch--achieved by varying the effective electron mass ($\Delta m_e = m_{e, \rm early} - m_{e,0}$ with $m_{e,0} = 511 \,\rm keV$)--provided the best performance across multiple evaluation metrics. Given the aforementioned degeneracy between $A_{\rm lens}$ and $\Omega_K$, this motivates us to extend the $\Delta m_e + \Omega_K$ scenario by allowing $A_{\rm lens}$ to vary as well. Our goal is to determine whether this extended model can offer a reasonable explanation for the lensing amplitude anomaly while simultaneously addressing the other cosmological tensions.

The paper is organized as follows: In Section~\ref{sec:Methods}, we describe the physical effects of varying $m_e$, $\Omega_K$, and $A_{\rm lens}$ on the CMB and other cosmological observables, and outline our analysis methodology. In Section~\ref{sec:Results}, we present the Bayesian parameter estimation results. We conclude with a summary and discussion in Section~\ref{sec:Conclu}.

\section{Methods}\label{sec:Methods}
\subsection{The role of varying constants}
In the early Universe, a simple description of the time-variation of $m_e$ can be constructed by assuming an instantaneous transition at some specific epoch. The electron mass is a fundamental constant that cannot be derived from first principles and is determined only experimentally; consequently, its potential variation over space or time has been widely explored in theoretical frameworks beyond $\Lambda$CDM \cite{2003Ap&SS.283..445D, 2003PhRvD..67a5009D, 2004NuPhB.677..471D,2022PhRvD.105j3536S}. The dominant impact of a varying $m_e$ on the CMB comes through modifications to the recombination history. The dependence of hydrogen and helium energy levels, atomic transition rates, and photoionization/recombination rates on $m_e$ has been detailed in prior works \cite{PhysRevD.82.063521, PhysRevD.83.043513}:
\begin{align}
\mathcal{A}_{2s}, \, \mathcal{A}_{2p} & \propto m_e^{-2} \qquad \qquad
\mathcal{B}_{2s}, \, \mathcal{B}_{2p}, \, \mathcal{R}_{2p2s}, \, \mathcal{R}_{2s2p} \propto m_e \qquad \qquad
\Lambda_{2s,1s}  \propto m_e \\
\sigma_T &\propto m_e^{-2} \qquad \qquad \qquad \qquad  T_{\rm eff} \propto m_e^{-1}
\end{align}
Here, $\mathcal{A}_{2s}$ and $\mathcal{A}_{2p}$ are the effective recombination coefficients to the lowest excited state ($n=2$), depending only on the matter and radiation temperatures. $\mathcal{B}_{2s}$ and $\mathcal{B}_{2p}$ represent the effective photoionization rates that describe hydrogen ionization by thermal CMB photons. $\mathcal{R}_{2p2s}$ and $\mathcal{R}_{2s2p}$ denote the effective transition rates between the $2s$ and $2p$ states via higher excitations. $\Lambda_{2s,1s}$ is the spontaneous two-photon decay rate from $2s$ to $1s$. The Thomson scattering cross section is $\sigma_T$, and $T_{\rm eff}$ is an effective temperature introduced to rescale the recombination and photoionization rates.

\begin{figure*}[htbp]
\centering
\includegraphics[width=1.\textwidth]{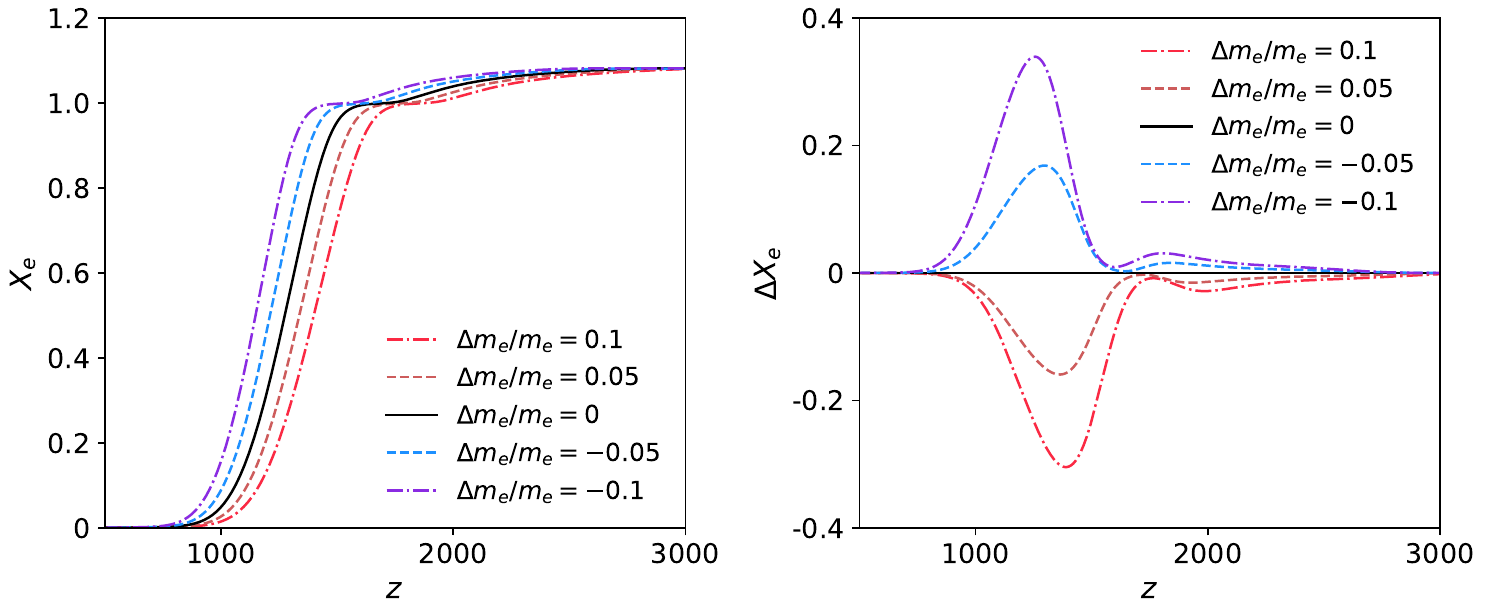}
\caption{Ionization history for different values of $\Delta m_e/m_e$. The left panel shows the evolution of $X_e$, while the right panel shows relative deviations with respect to the standard case. All curves are obtained using \texttt{HYREC-2}. \label{fig:Xe_evo}}
\end{figure*}

\begin{figure*}[htbp]
\centering
\includegraphics[width=1.\textwidth]{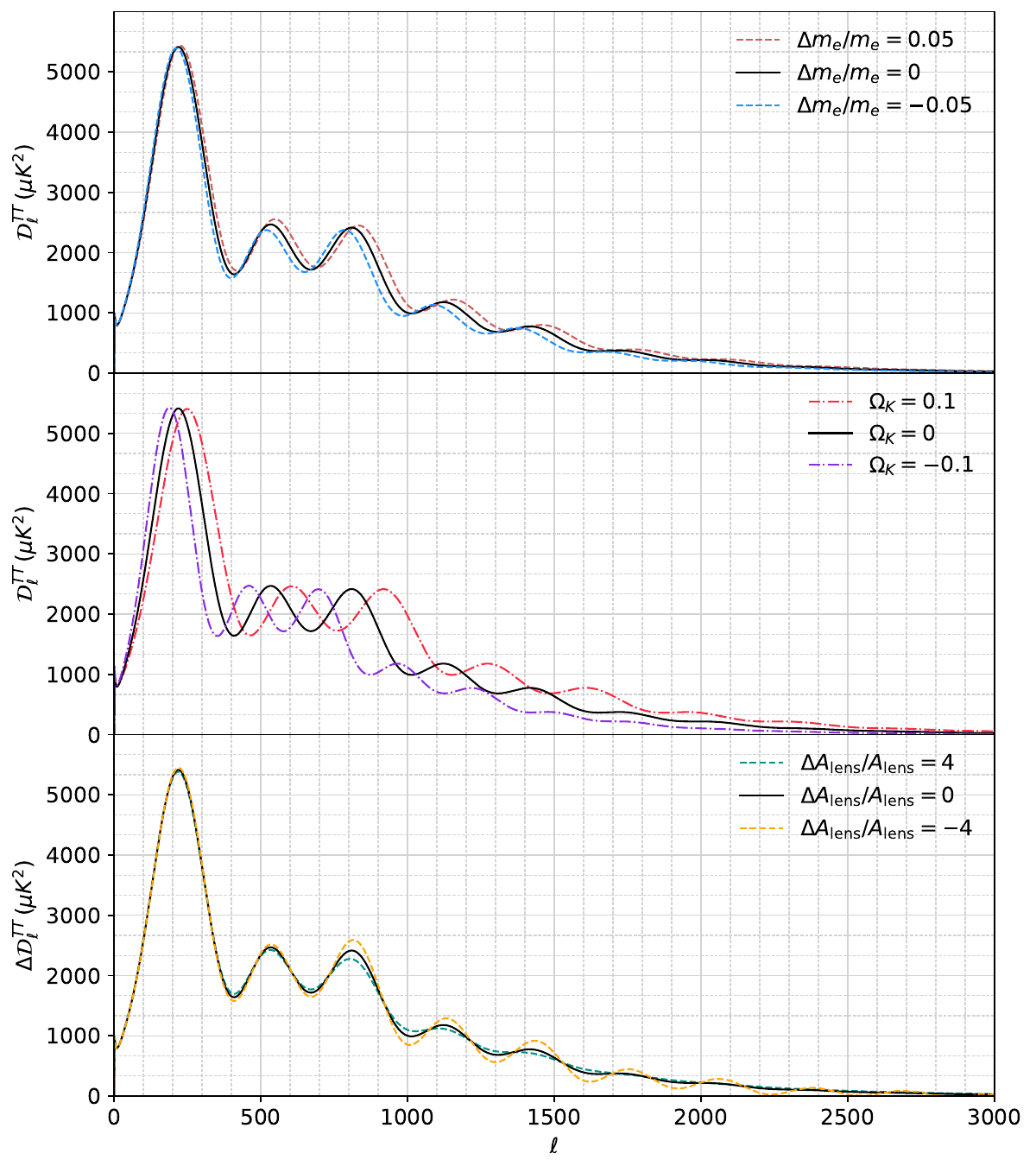}
\caption{The CMB temperature power spectra and their variations for different values of $m_e$, $\Omega_K$, and $A_{\rm lens}$, calculated by the cosmological Boltzmann code \texttt{CAMB} with the recombination code \texttt{HYREC-2}. \label{fig:TT_evo}}
\end{figure*}

Defining the free electron fraction as $X_e = n_e/n_{\rm H}$, its evolution obeys
\begin{equation}
\dot{X_e} = \sum_{\ell=s,p}(X_{2\ell}\mathcal{B}_{2\ell} - n_{\rm H}X_e^2\mathcal{A}_{2\ell}),
\end{equation}
where $n_{\rm H}$ is the total hydrogen number density, and $X_{2s}$ and $X_{2p}$ denote the fractional abundances of hydrogen in the $2s$ and $2p$ excited states. The evolution of $X_e$ for several values of $\Delta m_e/m_e$ is shown in \autoref{fig:Xe_evo}, computed using the recombination code \texttt{HYREC-2}\footnote{\url{ https://github.com/nanoomlee/HYREC-2}}. The corresponding modifications to the CMB temperature power spectrum for constant shifts in $m_e$ are shown in the top panel of \autoref{fig:TT_evo}. As $m_e$ increases, the heights of the acoustic peaks beyond the first peak are slightly enhanced. At the same time, the positions of the peaks shift toward higher multipoles $\ell$, since a larger electron mass leads to earlier recombination and thus pushes the last scattering surface to higher redshifts.

In a $\Lambda$CDM + $\Delta m_e$ scenario, once $m_e$ is varied there are no additional degrees of freedom to modify the late-time geometry. Consequently, resolving the $H_0$ tension through changes in $m_e$ alone proves difficult. Using the sound horizon at the recombination epoch and the relative scale of the Silk damping scale, ref.~\cite{PhysRevD.103.083507} proposed that under the $\Lambda$CDM late-time expansion history, the relation between the reduced Hubble constant $h$ and the scale factor at recombination $a_*$ is $\ln(h / h_{\rm fid})\approx -3.23 \ln(a_* / a_{*,\rm fid})$, where ``fid" denotes the fiducial $\Lambda$CDM cosmology. Considering the effect of $m_e$ on the CMB at recombination, $a_*$ is inversely proportional to $m_e$, following $\ln(m_e /m_{e,\rm fid}) \approx - \ln{(a_* /a_{*,\rm fid})}$. When both $m_e$ and $\Omega_K$ are allowed to vary, the relations $\ln(h /h_{\rm fid}) \approx 3.23 \ln{(m_e /m_{e,\rm fid})}$ and $\Omega_K \approx -0.125 \ln{(m_e /m_{e,\rm fid})}$ emerge \citep{PhysRevD.103.083507, 2025JCAP...03..004S}. Furthermore, the aforementioned modifications influence both the expansion history and the geometric distance, which can be constrained by BAO and SNe Ia observations. The primary observables constrained by BAO are the ratios $D_{\rm M}/r_{\rm d}$ and $D_{\rm H}/r_{\rm d}$ \citep{2013PhR...530...87W}. Here, $D_{\rm M}$ denotes the comoving angular diameter distance, defined as
\begin{displaymath}
D_{\rm M}(z) = \left\{ \begin{array}{ll}
\frac{c}{H_0\sqrt{\Omega_K}}\sinh{\big[\sqrt{\Omega_K} \int^z_0 \frac{{\rm d} z'}{H(z')/H_0}\big]}, & \Omega_K>0,\\
\frac{c}{H_0}\int^z_0 \frac{{\rm d} z'}{H(z')/H_0}, & \Omega_K=0,\\
\frac{c}{H_0\sqrt{|\Omega_K|}}\sin{\big[\sqrt{|\Omega_K|} \int^z_0 \frac{{\rm d} z'}{H(z')/H_0}\big]}, & \Omega_K<0.
\end{array} \right.
\end{displaymath}
The Hubble distance is defined as $D_{\rm H}(z) = c/H(z)$. The pre-recombination sound horizon (before photon decoupling), corresponding to the maximum distance acoustic waves could travel in the primordial plasma, is
\begin{equation}
r_{\rm d} = \int^\infty_{z_{\rm d}} \frac{c_s {\rm d}z}{H(z)},
\end{equation}
where $c_s$ is the pre-recombination sound speed and $z_{\rm d}$ is the redshift at drag epoch. For a variation $\Delta m_e$, the sound horizon shifts as $\ln(r_{\rm d}/r_{\rm d,fid})=-\ln(m_e/m_{e,\rm fid})$ \cite{2025JCAP...03..004S}, demonstrating that BAO data alone can not provide strong constraints on $m_e$. Furthermore, the acoustic peaks in the CMB power spectrum encode the physics of the recombination. The angular scale of these peaks measures the ratio $D_{\rm M}(z_*)/r_*$, with
\begin{equation}
r_{\rm *} = \int^\infty_{z_{\rm *}} \frac{c_s {\rm d}z}{H(z)},
\end{equation}
where $z_*$ is the redshift of recombination when photons decoupled with the baryons and $r_*$ is the comoving sound horizon at the end of recombination.

For the remaining extended parameters $\Omega_K$ and $A_{\rm lens}$, their effects on the CMB temperature angular power spectra are illustrated in \autoref{fig:TT_evo}. A larger positive curvature (open Universe) shifts the acoustic peaks to smaller angular scales, showing a similar trend to the effect of an increased $m_e$. As a phenomenological extension, the lensing amplitude $A_{\rm lens}$ is defined as a scaling factor affecting the lensing potential power spectrum and controlling the amount of smoothing of the peaks: $C^{\Psi}_{\ell} \to A_{\rm lens}C^{\Psi}_{\ell}$ \citep{2008PhRvD..77l3531C}, representing the amplitude of the lensing power relative to the physical value. Setting $A_{\rm lens}=0$ corresponds to neglecting CMB lensing entirely, while $A_{\rm lens}=1$ recovers the standard $\Lambda$CDM prediction.

\subsection{Methodology and data sets}
We perform Bayesian parameter estimation for several cosmological models, extending the baseline $\Lambda$CDM scenario. The extensions considered include a time-varying electron mass ($\Delta m_e$), a non-flat Universe ($\Omega_K$), an anomalous CMB lensing amplitude ($A_{\rm lens}$), and a dynamical dark energy model with CPL parameterization ($w_0,w_a$). The baseline $\Lambda$CDM model itself is specified by six primary parameters: the Hubble constant $H_0$, the cold dark matter density $\omega_c$, the baryon density $\omega_b$, the reionization optical depth $\tau$, the amplitude of the primordial scalar power spectrum $A_s$, and the scalar spectral index $n_s$. 

To balance the efficiency and accuracy, we employ a nested sampling algorithm using the \texttt{nessai} sampler \cite{2021PhRvD.103j3006W, 2024ascl.soft05002W}. We set the number of live points to $800$ and terminate the sampling once the fractional change in the log-evidence satisfies $\Delta \ln Z < 0.1$ \cite{Skilling2007}. Likelihood functions are computed with cosmological Boltzmann code \texttt{CAMB}\footnote{\url{https://github.com/cmbant/CAMB}} \cite{2000ApJ...538..473L, 2012JCAP...04..027H}, the recombination code \texttt{HYREC-2} \cite{PhysRevD.83.043513, 2020PhRvD.102h3517L}, and the cosmological inference framework \texttt{Cobaya}\cite{2019ascl.soft10019T, 2021JCAP...05..057T}.

In addition to CMB observations, BAO measurements provide extra constraints on $m_e$ and $\Omega_K$. Late-time cosmological probes such as SNe~Ia and cosmic chronometers can also improve sensitivity to spatial curvature. The cosmological data sets used in this analysis are as follows:

(1) {\bf CMB}. Two scenarios are considered. The first ensemble of CMB power spectra combines Planck and ACT DR6 observations. The Planck data set includes the high-$\ell$ TT power spectrum at $\ell<1000$, TE/EE power spectrum at $\ell<600$ data, and the low-$\ell$ TT power spectrum from the Planck PR3 likelihood \cite{2020A&A...641A...5P}, along with the low-$\ell$ polarization in EE spectrum from the LoLLiPoP likelihood \cite{2024A&A...682A..37T}. The ACT DR6 contributes the high-$\ell$ TT power spectrum at $\ell>1000$ and TE/EE power spectrum at $\ell>600$ \cite{2025arXiv250314452L}. 
The second scenario relies solely on Planck observations, where the high-$\ell$ data are replaced by the TT/TE/EE power spectrum from the Planck PR4 maps, including the NPIPE \citep{2022MNRAS.517.4620R} and HiLLiPoP \citep{2024A&A...682A..37T} likelihoods. 

(2) {\bf CMB lensing}. The lensing likelihood includes the Planck PR4 lensing reconstruction \cite{2022JCAP...09..039C} together with the ACT DR6 lensing map \cite{2024ApJ...962..113M, 2024ApJ...966..138M, 2025arXiv250708798E}, following the recommendation of ref.~\cite{2024ApJ...976L..11R}. 

(3) {\bf BAO}. The analysis incorporates the latest BAO measurements from DESI DR2 \cite{2025PhRvD.112h3515A}, including the distance ratios $D_{\rm V}/r_{\rm d}$, $D_{\rm M}/r_{\rm d}$, and $D_{\rm H}/r_{\rm d}$ at effective redshifts $z_{\rm eff}$. These data span $0.295<z<2.33$ using multiple tracers: the bright galaxy sample, Luminous red galaxies, emission line galaxies, quasars and Ly$\alpha$ forest.

(4) {\bf SNe Ia}. Two separate samples are adopted. The Pantheon+ compilation \cite{2022ApJ...938..110B, 2022ApJ...938..113S} consists of 1701 light curves of 1550 distinct SNe Ia covering $0.001<z<2.26$. The full 5 yr of the DES SN (DES Y5) sample \cite{2024ApJ...973L..14D} includes 1635 SNe Ia in the redshift range $0.10<z<1.13$.

(5) {\bf Cosmic Chronometers (CC)}. The CC technique determines $H(z)$ directly from the differential age evolution of galaxies, $H(z) = -\frac{1}{1+z} \frac{{\rm d}z}{{\rm d}t}$, without assuming a cosmological model \cite{2002ApJ...573...37J}. A compilation of 32 measurements in the range $0.07<z<1.965$ is used \cite{2022LRR....25....6M, 2024arXiv241201994M}, ensuring no overlap between data points obtained from different surveys by the same method.

\begin{figure*}[htbp]
\centering
\includegraphics[width=0.6\textwidth]{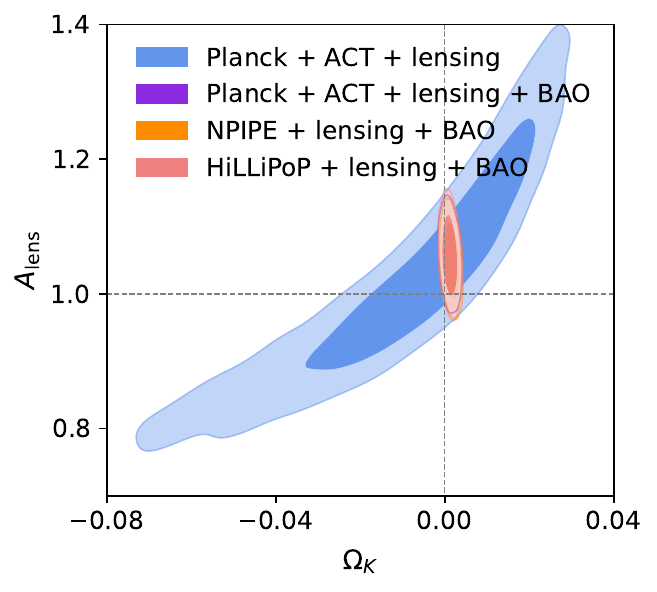}
\caption{Degeneracy between $\Omega_K$ and $A_{\rm lens}$. The blue region shows the posterior result from the combination of the TT/TE/EE power spectra (Planck PR3 + ACT, NPIPE, and HiLLiPoP) and the lensing maps (Planck PR4 + ACT DR6). The pink region further includes BAO measurements from DESI DR2. The dashed grey lines represent the expectations of $\Lambda$CDM model. All contours show 68\% and 95\% credible intervals.\label{fig:omega_k_A}}
\end{figure*}

\section{Results}\label{sec:Results}
Before discussing the cosmological tensions in detail, we first examine the flatness of the universe and the degeneracy between $\Omega_K$ and $A_{\rm lens}$. The extended parameters are $\Omega_K$ and $A_{\rm lens}$ only. Using CMB and CMB lensing data alone, the trend of the degeneracy shown in \autoref{fig:omega_k_A} is consistent with earlier findings in ref.~\cite{2020NatAs...4..196D}. In contrast to that work, our analysis uses the combined CMB power spectra and the joint lensing reconstruction from Planck and ACT DR6. We find no evidence for a closed Universe or any apparent lensing amplitude anomaly. Once BAO measurements are included, the degeneracy between curvature and lensing amplitude is broken. Three CMB power spectrum likelihoods yield consistent results. The combined CMB, lensing, and BAO data yield tight constraints, i.e., $\Omega_K = 0.0013\pm{0.0012} \, (0.0017^{+0.0011}_{-0.0012})$ and $A_{\rm lens} =1.055\pm{0.035} \, (1.048^{+0.037}_{-0.036})$ for Planck PR3 (PR4 NPIPE) scenario, respectively, suggesting a mild preference for an open Universe.

\begin{figure*}[htbp]
\centering
\subfigure{
	    \hspace{-6mm}
		\includegraphics[width=0.49\textwidth]{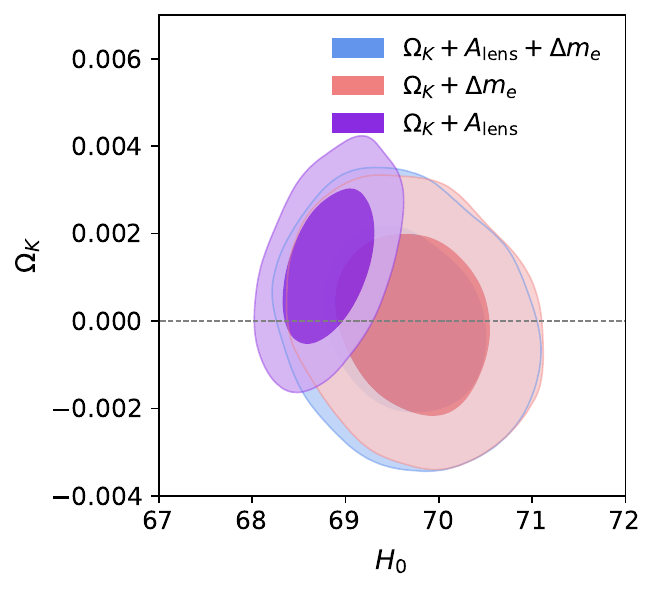}
		\hspace{2mm}
		\includegraphics[width=0.49\textwidth]{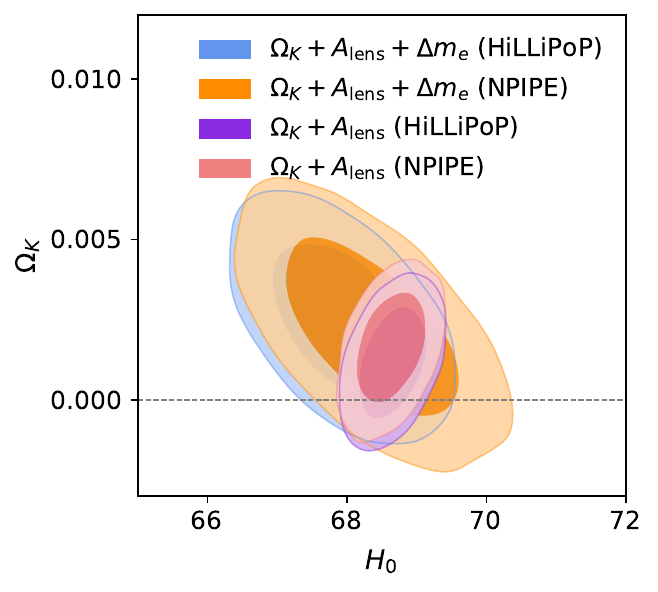}}
\caption{Posterior distributions of $\Omega_K$ and $H_0$ in different extended models. In the left panel, he combined data sets include the TT/TE/EE power spectra (Planck and ACT), the lensing maps (Planck PR4 + ACT DR6), and the BAO measurements (DESI DR2). $\Omega_K+A_{\rm lens}+ \Delta m_e$, $\Omega_K+ \Delta m_e$, and $\Omega_K+A_{\rm lens}$ are shown in blue, pink, and purple regions, respectively. The right panel shows the results using the lensing maps, BAO measurements and the PR4 NPIPE or HiLLiPoP likelihoods. The results of the three scenarios The dashed grey line indicates the expectation of $\Lambda$CDM model. The y-axis ranges differ between these two panels and \autoref{fig:omega_k_A}}. All contours show 68\% and 95\% credible intervals.\label{fig:omega_k}
\end{figure*}

In \autoref{fig:omega_k}, we show the posterior distributions in the $H_0 - \Omega_K$ plane for various extended models. The results are derived from the CMB power spectra (including the NPIPE, HiLLiPoP, and the combined Planck PR3 and ACT likelihoods), the lensing maps from Planck PR4 and ACT DR6, and the BAO measurements from DESI DR2. All of the $\Omega_K + A_{\rm lens}$ cases present positive curvatures with $\Omega_K=0.0013 \pm 0.0012$, $\Omega_K=0.0012 \pm 0.0011$ and $\Omega_K=0.0017^{+0.0011}_{-0.0012}$ for Planck PR3, HiLLiPoP, and NPIPE scenarios, suggesting a preference for an open Universe. 
In the left panel, compared to the $\Omega_K + A_{\rm lens}$ scenario, allowing a varying electron mass shifts the inferred $H_0$ from $68.81^{+0.33}_{-0.31}$ to $69.61^{+0.60}_{-0.55} \, \rm km \, s^{-1} \, kpc^{-1}$, relieving the Hubble tension from $4.1\sigma$ to $3.1\sigma$ \footnote{The tension refers to the Gaussian tension with ref. \cite{2022ApJ...934L...7R}}, while a flat Universe remains preferred. In the $\Omega_K+A_{\rm lens}+\Delta m_e$ case, we find $\Delta m_e/m_e =0.011^{+0.007}_{-0.006}$ and $A_{\rm lens}=1.03 \pm 0.04$. In the $\Omega_K + \Delta m_e$ model (without $A_{\rm lens}$), the electron mass shift remains significant which is $\Delta m_e/m_e = 0.013\pm0.006$. The contours of  $\Omega_K+A_{\rm lens}+\Delta m_e$ and $\Omega_K + \Delta m_e$ cases remain nearly identical, suggesting an orthogonality between the $A_{\rm lens}$ and $\Omega_K + \Delta m_e$. When using the Planck PR4 likelihoods, both HiLLiPoP and NPIPE provide consistent results, as shown in the right panel. A positive curvature is obtained in each case, even when allowing for a varying electron mass, which differs from the results based on the Planck PR3 likelihoods. Moreover, $H_0$ does not increase when introducing a varying electron mass, i.e. from $68.62^{+0.33}_{-0.32}$ to $68.34^{+0.80}_{-0.82} \, \rm km \, s^{-1} \, kpc^{-1}$ using the NPIPE likelihood. The corresponding Hubble tension remains at the $4.2\sigma-3.8\sigma$ level. This behavior arises because the PR4 likelihoods yield smaller electron mass, with $\Delta m_e/m_e =-0.0036^{+0.0095}_{-0.0099}$ and $-0.0095^{+0.0078}_{-0.0079}$ for the NPIPE and HiLLiPoP likelihoods, respectively.

\begin{figure*}[htbp]
\centering
\includegraphics[width=1.\textwidth]{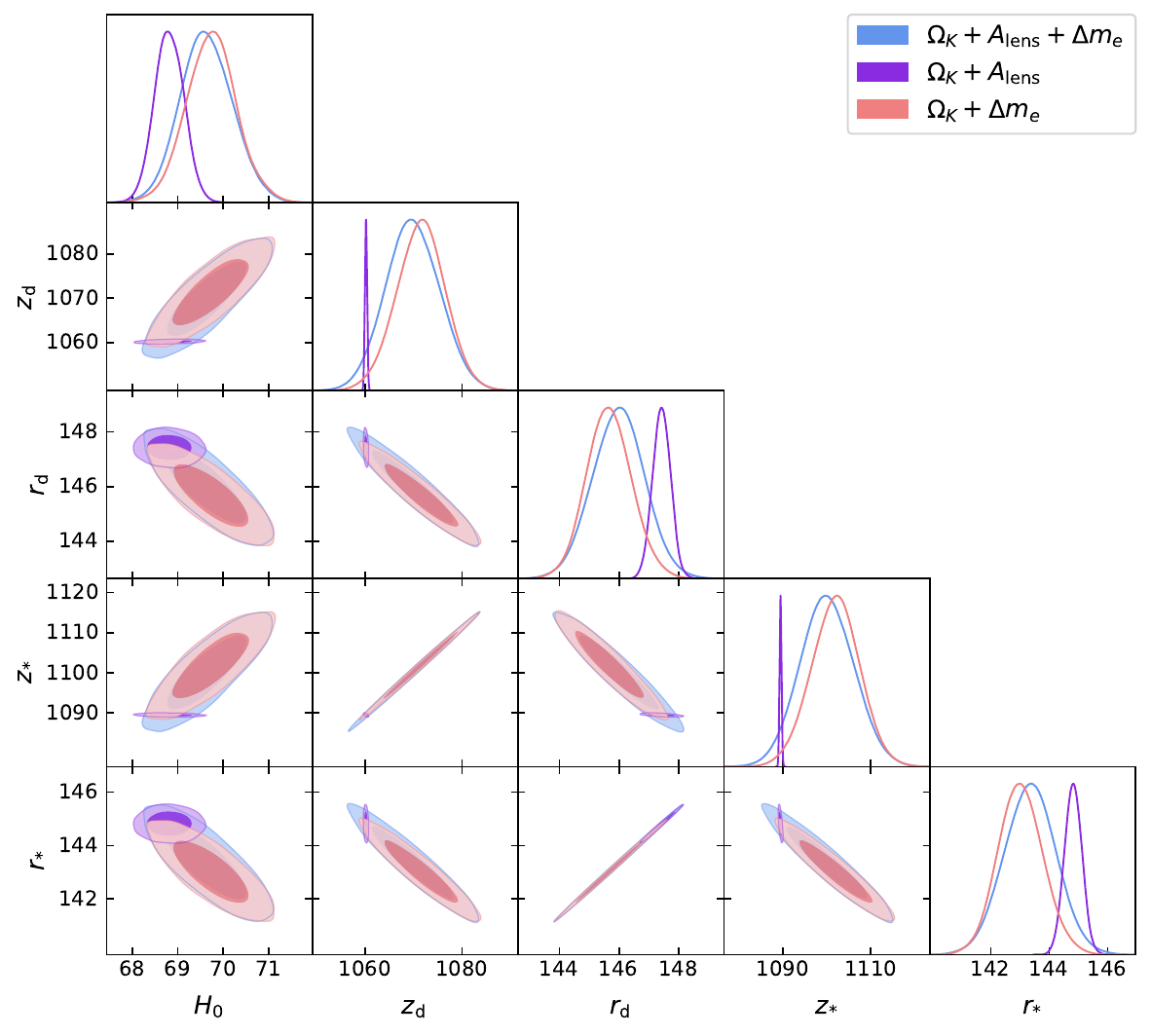}
\caption{Posterior distributions of $H_0$, $z_{\rm d}$, $r_{\rm d}$, $z_*$, and $r_*$ in different extended models. The definitions of the color regions are same with those in the left panel of \autoref{fig:omega_k}. All contours show 68\% and 95\% credible intervals.\label{fig:rdrag}}
\end{figure*}

Since the electron mass directly affects the sound horizon at recombination and at the drag epoch, we further examine the distributions of $r_{\rm d}$, $z_d$, $r_*$, and $z_*$ in \autoref{fig:rdrag}, corresponding to the left panel in \autoref{fig:omega_k}. A strong degeneracy between $H_0$ and these four parameters is evident, and an clear degeneracy between $r_*$ and $r_d$ still remains in the extended cosmological models. Compared with the $\Omega_K + A_{\rm lens}$ scenario, the $\Omega_K + \Delta m_e$ and $\Omega_K + A_{\rm lens} + \Delta m_e$ scenarios predict a higher electron mass, an advance in the recombination epoch, a reduction in the sound horizon, and thereby drive $H_0$ toward higher values. 
For the $\Omega_K + \Delta m_e$ scenario compared with the $\Lambda$CDM model, we obtain $\ln(m_e/m_{e, \rm fid}) \approx 0.013$, $\ln(h_{\rm}/h_{\rm fid}) \approx 0.018$, $\ln(r_{\rm d}/r_{\rm d,fid})\approx -0.013$, and $\ln(a_*/a_{*, \rm fid})\approx -0.012$, consistent with the relations $\ln(m_e/m_{e, \rm fid}) \approx -\ln(a_*/a_{*, \rm fid})$ and $\ln(m_e/m_{e, \rm fid}) \approx -\ln(r_{\rm d}/r_{\rm d,fid})$ discussed in Section~\ref{sec:Methods}. However, the relation between $h$ and $m_e$ gives $\ln(h_{\rm}/h_{\rm fid}) \approx 1.40 \ln(m_e/m_{e, \rm fid})$, which is smaller than the analytical expectation. This occurs because, when additional parameters such as $\Omega_K$, $A_{\rm lens}$, or $\Delta m_e$ are allowed to vary, the relation of $\ln(h / h_{\rm fid})\approx -3.23 \ln(a_* / a_{*,\rm fid})$ (as refereed in Section~\ref{sec:Methods}) becomes model-dependent, thereby reducing the effective slope between $h$ and $m_e$.

\begin{figure*}[htbp]
\centering
\includegraphics[width=0.7\textwidth]{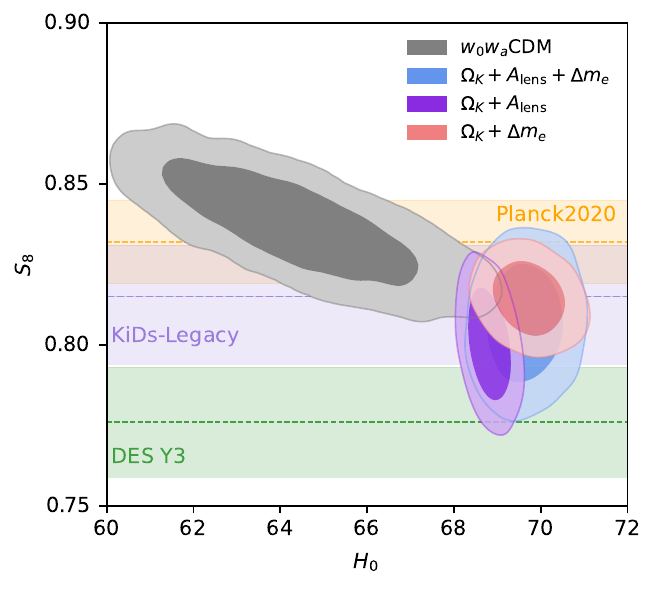}
\includegraphics[width=0.7\textwidth]{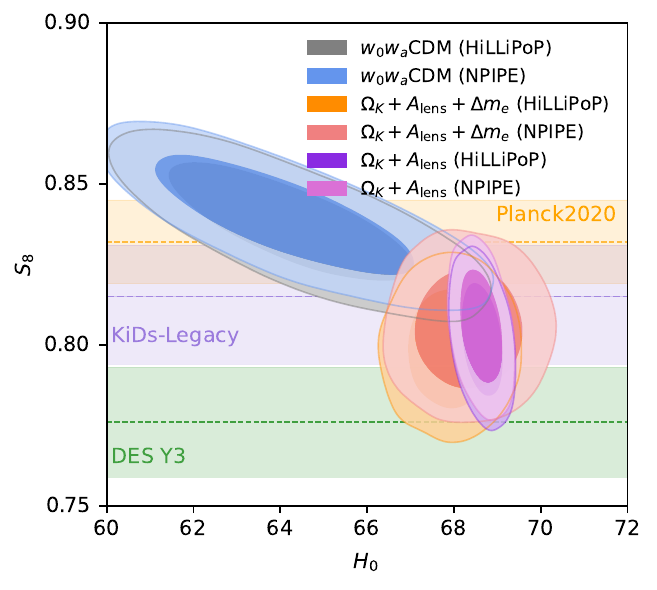}
\caption{Posterior constraints in the $H_0$-$S_8$ plane for different cosmological models, consistent with the definitions in \autoref{fig:omega_k}. The top panel uses the combination of Planck PR3 and ACT likelihoods, and the bottom panel uses the Planck PR4 NPIPE and HiLLiPoP likelihoods. The grey region corresponds to the $w_0 w_a$CDM model. The yellow and green regions indicate the Planck 2018 results ($S_8=0.832\pm0.013$) \cite{2020A&A...641A...6P}, DES Y3 results ($S_8 = 0.776\pm0.017$) \cite{2022PhRvD.105b3520A}, and latest KiDs-Legacy results ($S_8 = 0.815^{+0.016}_{-0.021}$) \cite{2025arXiv250319441W} respectively. All contours show 68\% and 95\% credible intervals.\label{fig:S8_H0}}
\end{figure*}

We next examine the impact on the $S_8$ tension when extending the cosmological model with $w_0$ and $w_a$, and with various combinations of $\Omega_K$, $A_{\rm lens}$, and $m_e$. \autoref{fig:S8_H0} shows constraints in the $H_0$-$S_8$ plane, based on the combination of the CMB power spectra from Planck and ACT, the Planck PR4 likelihoods NPIPE and HiLLiPoP, the lensing maps from Planck PR4 and ACT DR6, and the BAO measurements from DESI DR2. In the late Universe, the amplitude of matter clustering is quantified by $S_8 \equiv \sigma_8 \sqrt{ \big( \frac{\Omega_{\rm m}}{0.3} \big)}$, where $\sigma_8$ is the rms amplitude of matter perturbations on $8h^{-1}$ Mpc scales. The top panel presents the results for the Planck PR3 + ACT scenario. We find that in the $w_0 w_a$CDM model, the Hubble tension is exacerbated (yielding $H_0=64.4^{+2.0}_{-1.9} \rm \, km \, s^{-1} \, Mpc^{-1}$), while the inferred $S_8=0.837^{+0.012}_{-0.013}$ remains consistent with Planck 2018 results. By contrast, the extensions involving $\Omega_K$, $A_{\rm lens}$, and $m_e$ relieve both the Hubble tension and $S_8$ tension because of the broadened uncertainties. For instance, the $\Omega_K+A_{\rm lens}+\Delta m_e$ and $\Omega_K+A_{\rm lens}$ scenarios yield $S_8 = 0.807\pm0.012$ ($\sigma_8 = 0.812\pm0.010$) and $0.800\pm0.011$ ($\sigma_8 = 0.802^{+0.008}_{-0.009}$), respectively. The broadened uncertainties in $S_8$ originate from the uncertainties in both $\Omega_{\rm m}$ and $\sigma_8$. Using the approximation $\sigma_{S_8 (\Omega_{\rm m})}\sim |\frac{\partial{S_8}}{\partial{\Omega_{\rm m}}}| \sigma_{\Omega_{\rm m}}$, we find that the contribution from $\Omega_m$ accounts for roughly a half of the total $S_8$ uncertainty in both $\Omega_K+A_{\rm lens}+\Delta m_e$ and $\Omega_K+ A_{\rm lens}$ scenarios. Nevertheless, the $S_8$ tension remains in the $\Omega_K + \Delta m_e$ case, with $S_8 = 0.814^{+0.008}_{-0.007}$, leaving a $\sim2.0 \sigma$ tension relative to the DES results \citep{2022PhRvD.105b3520A}, but only a $\sim 0.08 \sigma$ relative to the KiDs-Legacy results \citep{2025arXiv250319441W}. The differing results between the various combinations of $\Omega_K$, $A_{\rm lens}$, an $\Delta m_e$ show that the orthogonality between the $A_{\rm lens}$ and $\Delta m_e$ produces a complementary effect in relieving the $S_8$ tension, with $\Omega_K$ gluing both neatly together. These different results of $S_8$ are mainly driven by the change in $\Omega_m$: the $w_0 w_a$CDM model predicts a larger matter fraction with $\Omega_m = 0.343\pm0.022$, whereas the $\Omega_K + A_{\rm lens} + \Delta m_e$ model has $\Omega_m = 0.296\pm0.004$. The bottom panel shows the constraints derived using the Planck PR4 likelihoods. The variation trend is similar to that shown in the left panel. However, only the $S_8$ tension is alleviated. For the $\Omega_K+A_{\rm lens}+\Delta m_e$ ($\Omega_K+ A_{\rm lens}$) scenario, we obtain $S_8 = 0.805\pm0.012$ ($0.800^{+0.013}_{-0.012}$) and $0.801^{+0.012}_{-0.011}$ ($0.799\pm0.012$) when using the NPIPE and HiLLiPoP likelihoods. The corresponding tensions compared to DES are $\sim 1.4\sigma$ and $\sim 1.2 \sigma$, respectively. In all $\Omega_K+A_{\rm lens}+\Delta m_e$ and $\Omega_K+ A_{\rm lens}$ cases regardless of the release version of the Planck likelihoods, the $S_8$ constraints show consistency with the latest KiDs-Legacy measurements \cite{2025arXiv250319441W}.

\begin{figure*}[htbp]
\centering
\includegraphics[width=0.7\textwidth]{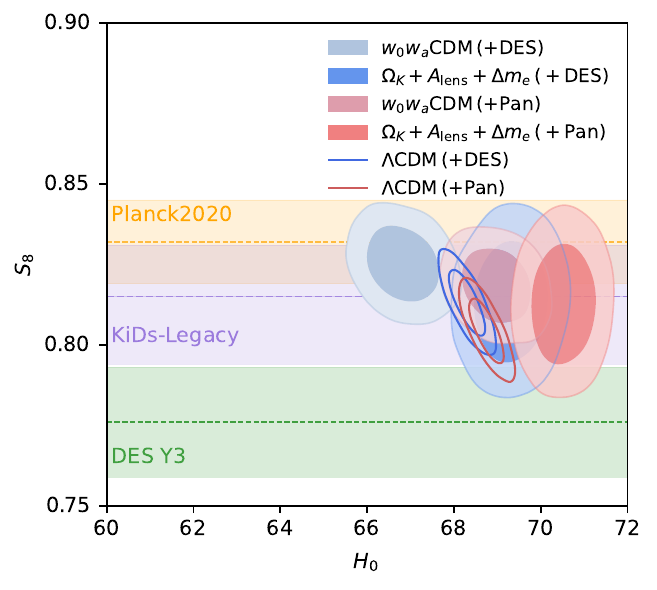}
\caption{Posterior constraints in the $H_0$-$S_8$ plane with the addition of local probes. The combined data sets include the CMB power spectra from Planck and ACT, the lensing maps from Planck PR4 and ACT DR6, the BAO measurements from DESI DR2, $H(z)$ measurements from CC data sets, and the SNe Ia samples. The red and blue regions represent the constraints with Pantheon+ or DES Y5, respectively. The solid lines indicate the expectations of the $\Lambda$CDM model. All contours show $68\%$ and $95\%$ credible intervals.\label{fig:S8_H0_local}}
\end{figure*}

Furthermore, we test the robustness of these conclusions by including local cosmological probes. \autoref{fig:S8_H0_local} shows the results after adding cosmic chronometer $H(z)$ data and SNe Ia samples (Pantheon+ or DES Y5). Within the $\Omega_K+A_{\rm lens}+\Delta m_e$ scenario, we find $H_0 = 69.30^{+0.55}_{-0.57}$ and $70.51\pm0.49\,\rm km \,s^{-1}\, Mpc^{-1}$ for DES Y5 and Pantheon+ data sets. The derived distributions of $S_8$ remain consistent with previous results, which are $S_8 = 0.813\pm0.012$ (DES Y5) and $S_8 = 0.812^{+0.012}_{-0.013}$ (Pantheon+).
Consequently, the Hubble tension is reduced to $\sim 3.4\sigma$ and $\sim 2.4 \sigma$, while the $S_8$ tension is reduced to $\sim 1.8\sigma$ and $\sim 1.7 \sigma$\footnote{The tension refers to the Gaussian tension with ref. \cite{2022PhRvD.105b3520A}}. Compared with the KiDs-Legacy measurements, the $S_8$ constraints are nearly consistent, corresponding to only $\sim 0.1\sigma$ level differences. Although the $w_0w_a$CDM model continues to predict a lower $H_0$ than the $\Lambda$CDM model, using different SNe Ia samples leads to noticeable discrepancies in the posteriors. Similar findings have been reported by recent analyses \cite{2024ApJ...976L..11R, 2025PhRvD.112b3508G, 2025arXiv250204212H}. As pointed out by ref.~\cite{2025MNRAS.538..875E}, these discrepancies may be partly due to a $\sim0.04$ mag offset between low- and high-redshift SNe when comparing DES Y5 to Pantheon+. A subsequent study found that this offset is largely attributable to differing bias-correction procedures \cite{2025MNRAS.541.2585V}.

A complete list of parameter constraints for all scenarios is provided in \autoref{sec:app}, with uncertainties quoted to two significant digits.

\section{Discussion and Conclusions}\label{sec:Conclu}
To evaluate the relative performance of the extended cosmological models, we computed the Bayes factor ($\ln \mathcal{B}$) and the Akaike Information Criterion (AIC) for each extended model $\mathcal{M}$ against the $\Lambda$CDM model, defined as
\begin{equation}
\begin{aligned}
\ln \mathcal{B} &= \ln \frac{ Z(\boldsymbol{y}\,|\,\mathcal{M})}{ Z(\boldsymbol{y}\,|\,\Lambda \mathrm{CDM}) }, \\
\Delta {\rm AIC} &= -2 \ln\!\left(\frac{L_{{\rm max}, \,\mathcal{M}}}{L_{{\rm max}, \,\Lambda \mathrm{CDM}}}\right) + 2(N_{\mathcal{M}} - N_{\Lambda \mathrm{CDM}}),
\end{aligned}
\end{equation}
where $\boldsymbol{y}$ denotes the data set, $Z$ is the Bayesian evidence of a given model, $L_{\rm max}$ is the maximum likelihood, and $N$ is the number of free parameters. The uncertainty of the Bayes factor $\sigma_{\ln{\mathcal{B}}}$ is estimated through error propagation as $\sigma_{\ln{\mathcal{B}}}^2 = \sigma_{\ln{Z_{\mathcal{M}}}}^2 + \sigma_{\ln{Z_{\rm \Lambda CDM}}}^2$.
A positive $\ln \mathcal{B}$ or a negative $\Delta {\rm AIC}$ indicates a statistical preference for the extended model, while both criteria penalize model complexity and over fitting.

Using the combined data set of CMB power spectra, CMB lensing reconstructions, BAO, CC, and SNe Ia, we found that most of the extended models yield $\ln \mathcal{B}<0$ and $\Delta {\rm AIC}<0$, with only three exceptions. This means that, in general, the Bayesian evidence disfavors these extensions, whereas the AIC indicates mild support. Specifically, the $w_0w_a$CDM model listed in \autoref{tab:2} and the $\Omega_K + \Delta m_e$ model listed in \autoref{tab:4} both present positive $\ln \mathcal{B}$ and negative $\Delta {\rm AIC}$, with $\ln \mathcal{B} = 3.29\pm0.28$ and $1.66\pm0.29$, and $\Delta {\rm AIC} = -14.01$ and $-17.42$, respectively, indicating a statistical preference for these models. In contrast, for the $w_0w_a$CDM model listed in \autoref{tab:4}, the Bayes factor $\ln \mathcal{B} = 0.60\pm0.27$ does not show a clear preference relative to the $\Lambda$CDM model. With respect to data that do not exhibit a large tension with CMB constraints, we find that these data (utilized in this work) are not yet sufficient to unambiguously determine whether the $\Lambda$CDM model or its extensions offer a better description of cosmological observations. It should be emphasized that the model preference may strongly depend on whether the combined data sets include measurements that are in tension with the standard $\Lambda$CDM framework. When priors from local measurements, such as the local $H_0$ and $S_8$, are incorporated, the $\Lambda$CDM model generally fails to perform a  good fit. In such cases, model extensions introducing additional degrees of freedom are proposed primarily to reconcile the tensions, rather than to improve the fit to datasets that are mutually consistent.

Our results show that the extensions involving $\Omega_K$, $A_{\rm lens}$, and $\Delta m_e$ when using the Planck PR3 likelihoods remain promising candidates for alleviating both the Hubble tension and the $S_8$ discrepancy. Only the $S_8$ tension can be relieved when using the Planck PR4 likelihoods instead.
When only considering a data set of CMB power spectra, CMB lensing reconstructions, BAO, but without local Universe priors, we obtain
\begin{equation}
\left. \begin{array}{l}
 { {H_0}   = 69.61^{+0.60}_{-0.55} \rm \, km \, s^{-1} \, Mpc^{-1}} \\
 {S_8   = 0.808\pm0.012} \\
 \end{array} \right\} 
 {\rm Planck \, PR3 + ACT + lensing + DESI \, DR2,} 
\end{equation}
along with a mild lensing amplitude anomaly $A_{\rm lens} = 1.030^{+0.039}_{-0.037}$ and an electron mass about $1\%$ higher than its standard value ($\Delta m_e/m_e = 0.0109^{+0.0068}_{-0.0066}$). The Hubble and $S_8$ tensions are reduced to approximately $3.3\sigma$ and $1.8\sigma$ significance, respectively. For the cases with PR4 likelihoods, we obtain
\begin{equation}
\left. \begin{array}{l}
 { {H_0}   = 68.34^{+0.80}_{-0.82} \rm \, km \, s^{-1} \, Mpc^{-1}} \\
 {S_8   = 0.800^{+0.013}_{-0.012}} \\
 \end{array} \right\} 
 {\rm NPIPE + lensing + DESI \, DR2,} 
\end{equation}
\begin{equation} 
 \left. \begin{array}{l}
 { {H_0}   = 67.94\pm0.067 \rm \, km \, s^{-1} \, Mpc^{-1}} \\
 {S_8   = 0.799\pm0.012} \\
 \end{array} \right\} 
 {\rm HiLLiPoP + lensing + DESI \, DR2,} 
\end{equation} 
still accompanied by a mild lensing amplitude anomaly ($A_{\rm lens} = 1.053^{+0.042}_{-0.040}$ and $1.075^{+0.044}_{-0.043}$), and by a smaller electron mass relative to the standard value ($\Delta m_e/m_e = -0.0036^{+0.0095}_{-0.0099}$ and $-0.0095^{+0.0078}_{-0.0079}$). Including the CC and Pantheon+ SNe data, we derive
\begin{equation}
\left. \begin{array}{l}
 { {H_0}   = 70.51\pm0.49 \rm \, km \, s^{-1} \, Mpc^{-1}} \\
 {S_8   = 0.813\pm0.012} \\
 \end{array} \right\} \begin{aligned} 
 &{\rm Planck + ACT + lensing + DESI \, DR2}\\
 &{\rm + CC + Pantheon+.} 
\end{aligned} 
\end{equation}
This scenario also prefers a non-standard electron mass, which is about $2\%$ higher than the standard value ($\Delta m_e/m_e = 0.020 \pm 0.0013$). Here, the Hubble and $S_8$ tensions are reduced to $2.4\sigma$ and $1.5\sigma$ significance, respectively. These results are consistent with several recent studies \cite{2024PhRvD.110h3501S,2024PDU....4601676T,2024JCAP...04..059K,2025JCAP...03..004S,2025arXiv250809025T,2025arXiv250620707C}. Meanwhile, the updated lensing amplitude $A_{\rm lens} = 1.008^{+0.036}_{-0.038}$ lies closer to the $\Lambda$CDM expectation. 
We further analyze the scenario for $\Delta m_e + \Omega_{\rm K}$ model and derive
\begin{equation}
\left. \begin{array}{l}
 { {H_0}   = 70.53^{+0.47}_{-0.46} \rm \, km \, s^{-1} \, Mpc^{-1}} \\
 {S_8   = 0.815^{+0.0073}_{-0.0072}} \\
 \end{array} \right\} \begin{aligned} 
 &{\rm Planck + ACT + lensing + DESI \, DR2}\\
 &{\rm + CC + Pantheon+.} 
\end{aligned} 
\end{equation}
The inferred Hubble constant remains in $2.4\sigma$ tension compared to the $\Omega_{K}+A_{\rm lens}+\Delta m_e$ case, while the $S_8$ tension is slightly enhanced to $2.1\sigma$ owing to the smaller uncertainty in $S_8$.
We find no evidence for a non-flat spatial curvature, in agreement with other recent analyses \cite{2021ApJ...908...84V,2023MNRAS.523.3406F,2023MNRAS.523.6360W,2024PhLB..85338699G,2024A&A...682A..37T} that used diverse methods and data sets. However, using the Planck HiLLiPoP and LoLLiPoP likelihoods, refs. \citep{2024ApJ...976L..11R, 2025ApJ...986L..31R} reported positive and negative significance of the lensing amplitude anomaly when including and excluding the weak lensing measurements. When using the high multipoles data alone, the lensing amplitude anomaly was also disfavored ~\citep{2025JCAP...02..069B}. Taken together, these findings suggest that while extensions to $\Lambda$CDM can modestly alleviate cosmological tensions, present observations do not yet provide decisive evidence for departures from the standard model.

Due to the cosmological constant tensions between the early- and late-Universe measurements, many studies have proposed models with evolving cosmological ``constants" (e.g., $m_e$, $\Omega_K$, $\Omega_m$, $H_0$, and $w$) to reconcile the discrepancies with observations. For example, ref.~\cite{PhysRevLett.130.161003} demonstrated that a time-varying electron mass $m_e(z)$ and fine-structure constant $\alpha(z)$ could address the Hubble tension and reduce the value of $S_8$ to match weak lensing measurements. The dark energy equation of state has also been parameterized in redshift bins, providing strong evidence for an evolving $w(z)$ \cite{2025PhRvD.111l3504P} and even a jointly evolving $H_0$ \citep{2025arXiv251014390W}. Moreover, non-parametric reconstructions of the dark energy equation of state have been employed to explain BAO observations \cite{2025arXiv250406118G, 2025MNRAS.540.2253A}, theoretically supported by a wide range of studies (i.e. refs \citep{2005PhLB..608..177G, 2007JHEP...10..071C, 2022PhRvD.106e5014Y, 2024JHEP...05..327Y, 2025arXiv250524732C}). Ref.~\cite{2025EPJC...85..286O} further reported indications of evolving cosmological parameters, identifying a possible redshift evolution in $\Omega_K$ using binned DES Y5 SNe Ia samples. Similarly, refs.~\cite{2022PhRvD.106d1301O,2024PDU....4401464O} found an increasing $\Omega_m$ accompanied by a decreasing $H_0$. Comparable evidence for an evolving $H_0$ has also been extensively discussed in refs.~\cite{2020PhRvD.102j3525K, 2021ApJ...912..150D, 2021PhRvD.103j3509K, 2023A&A...674A..45J, 2024EPJC...84..317M, 2024Univ...10..305A, 2025JHEAp..4800405D, 2025ApJ...979L..34J, 2025JCAP...03..026L, 2025arXiv250314743L, 2025PDU....4801847M, 2025MNRAS.536.3232M}.

Such extensive studies of cosmological frameworks across binned redshifts highlight the necessity of probing the Universe at different distances with diverse tracers and complementary methods. Beyond the traditional approaches, several novel astrophysical probes have recently been proposed to constrain the spatial curvature of the Universe (see refs. \cite{2022LRR....25....6M, 2025PDU....4901965D} for reviews). For instance, the dispersion measure of fast radio bursts (FRBs) provides an independent avenue for curvature measurements \citep{2025arXiv250601504F}, while the dust-scattering rings of gamma-ray bursts \citep{2025JCAP...06..032S} and the stochastic gravitational wave background \citep{2025JCAP...01..011C} offer additional and complementary constraints. Localized FRBs are particularly promising, as they can break the degeneracy between the dark energy equation of state and other cosmological parameters when combined with BAO measurements \citep{2014PhRvD..89j7303Z, 2025ApJ...981....9W}.

Looking ahead, forthcoming cosmological surveys will greatly enhance our ability to test extended models with varying fundamental constants and other cosmological parameters. In particular, next-generation CMB experiments such as CMB-S4, the Simons Observatory, and LiteBIRD \citep{2022arXiv220308024A, 2019BAAS...51g.147L, 2023PTEP.2023d2F01L}, together with high-precision BAO measurements from Euclid \citep{2025A&A...697A...1E} and LSST at the Rubin Observatory \citep{2019ApJ...873..111I}, will significantly improve the precision of cosmological parameters and potentially determine whether cosmological tensions originate from late-time or early-time new physics.

\appendix
\section{The results of parameter estimations for the standard and extended cosmological models}\label{sec:app}

This appendix presents the posterior distributions of the fundamental and extended cosmological parameters, along with derived quantities such as $z_*$, $z_{\rm d}$, $r_*$, $r_{\rm d}$, $\Omega_m$, and $S_8$, as well as relevant statistical metrics. In addition to the parameters of primary interest (including $H_0$, $S_8$, $w_0$, $w_a$, $A_{\rm lens}$, $\Omega_K$, and $\Delta m_e /m_e$), the uncertainties of the other parameters are quoted as the average of the upper and lower errorbars, i.e., $\sigma = \sqrt{\sigma^2_+/2 + \sigma^2_-/2}$.

\begin{sidewaystable*}[htbp]
\renewcommand{\arraystretch}{1.3} 
\begin{tabular}{l|ccccc}
\hline
\multicolumn{6}{c}{Planck 2018 + ACT DR6 + Lensing + DESI DR2}\\
\hline
Parameter &$\Lambda \rm CDM$ &$w_0 w_a \rm CDM$ &$A_{\rm lens} + \Omega_K$ &$\Delta m_e + \Omega_K$ &$A_{\rm lens} + \Delta m_e + \Omega_K$\\
\hline
$\Omega_b h^2$ &$0.02257\pm0.00010$ &$0.02250\pm0.00011$ &$0.02256\pm0.00011$ &$0.02256\pm0.00011$ &$0.02258\pm0.00011$\\
$\Omega_c h^2$ &$0.11724\pm0.00066$ &$0.11891\pm0.00092$ &$0.1180\pm0.0012$ &$0.1209\pm0.0015$ &$0.1202\pm0.0018$\\
$\tau$ &$0.0628\pm0.0062$ &$0.0590\pm0.0059$ &$0.579\pm0.006$ &$0.0577\pm0.0061$ &$0.0563\pm0.0063$\\
$\ln(10^{10} A_s)$ &$3.064\pm0.011$ &$3.049\pm0.011$ &$3.039\pm0.017$ &$3.048\pm0.011$ &$3.038\pm0.017$\\
$n_s$ &$0.9754\pm0.0035$ &$0.9715\pm0.0038$ &$0.9735\pm0.0042$ &$0.9621\pm0.0059$ &$0.9643\pm0.0066$\\
$H_0 \, [\rm km \,s^{-1} \, Mpc^{-1}]$ &$68.48^{+0.28}_{-0.27}$ &$64.36^{+1.98}_{-1.93}$ &$68.81^{+0.33}_{-0.31}$ &$69.74^{+0.52}_{-0.55}$ &$69.61^{+0.60}_{-0.55}$\\
$\Omega_K$ &$-$ &$-$ &$0.0013\pm0.0012$ &$0.0000\pm0.0013$ &$0.0000\pm0.0014$\\
$A_{\rm lens}$ &$-$ &$-$ &$1.055\pm0.035$ &$-$ &$1.030^{+0.039}_{-0.037}$\\
$\Delta m_e /m_e$ &$-$ &$-$ &$-$ &$0.0131^{+0.0058}_{-0.0059}$ &$0.0109^{+0.0066}_{-0.0062}$\\
$w_0$ &$-$ &$-0.51^{+0.22}_{-0.21}$ &$-$ &$-$ &$-$\\
$w_a$ &$-$ &$-1.41^{+0.58}_{-0.63}$ &$-$ &$-$ &$-$\\
\hline
$z_{*}$ &$1089.38\pm0.15$ &$1089.63\pm0.19$ &$1089.47\pm0.22$ &$1102.09\pm5.43$ &$1099.95\pm6.06$\\
$r_{\rm *}\, [\rm Mpc]$ &$145.00\pm0.18$ &$144.63\pm0.23$ &$144.82\pm0.30$ &$143.01\pm0.76$ &$143.36\pm0.88$\\
$z_{\rm d}$ &$1060.17\pm0.23$ &$1060.13\pm0.23$ &$1060.19\pm0.22$ &$1071.60\pm4.97$ &$1069.73\pm5.54$\\
$r_{\rm d}\, [\rm Mpc]$ &$147.61\pm0.20$ &$147.25\pm0.24$ &$147.43\pm0.31$ &$145.66\pm0.75$ &$146.00\pm0.87$\\
$\Omega_m$ &$0.2997\pm0.0036$ &$0.3433\pm0.2167$ &$0.2984\pm0.0038$ &$0.2965\pm0.0039$ &$0.2961\pm0.0040$\\
$S_8$ &$0.8091^{+0.0068}_{-0.0064}$ &$0.837^{+0.012}_{-0.013}$ &$0.800\pm0.011$ &$0.8144^{+0.0075}_{-0.0073}$ &$0.807\pm0.012$\\
\hline
$\ln(Z)$ &$-271.56\pm0.19$ &$-273.39\pm0.21$ &$-277.67\pm0.22$ &$-277.63\pm0.22$ &$-279.44\pm0.22$ \\
$\ln \mathcal{B}$ &$0$ &$-1.83\pm0.28$ &$-6.11\pm0.29$ &$-6.07\pm0.29$ &$-7.88\pm0.29$\\ 
max$\ln(L)$ &-237.61 &-234.45 &-235.46 &-234.10 &-233.46\\
$\Delta \rm AIC$ &$0$ &-2.32 &-0.28 &-3.02 &-2.30\\
\hline
\end{tabular}
\caption{68\% credible intervals for the $\Lambda$CDM and several extended models, using Planck and ACT CMB power spectra, in combination with Planck PR4 and ACT DR2 CMB lensing reconstruction and BAO measurements from DESI DR2. The top group of 11 rows are the base parameters, which are sampled in the Bayesian analysis with flat priors. The middle group lists derived parameters. The bottom group shows statistical metrics quantifying the fit efficiency.\label{tab:1}}
\end{sidewaystable*}

\begin{sidewaystable*}[htbp]
\renewcommand{\arraystretch}{1.3} 
\begin{tabular}{l|c@{\hspace{5.0em}}c@{\hspace{5.0em}}c@{\hspace{5.0em}}c}

\hline
\multicolumn{5}{c}{Planck 2018 + ACT DR6 + Lensing + DESI DR2 + CC + DES Y5}\\
\hline
Parameter &$\Lambda \rm CDM$ &$w_0 w_a \rm CDM$ &$\Delta m_e + \Omega_K$  &$A_{\rm lens} + \Delta m_e + \Omega_K$\\
\hline
$\Omega_b h^2$ &$0.02255\pm0.0010$ &$0.02252\pm0.00011$ &$0.02251\pm{0.00011}$ &$0.02252\pm0.00012$ \\
$\Omega_c h^2$ &$0.11771\pm0.00064$ &$0.11849\pm0.00088$ &$0.1213\pm{0.0015}$ &$0.1206\pm0.0018$ \\
$\tau$ &$0.0620\pm0.0061$ &$0.0560\pm0.0060$ &$0.0571\pm{0.0060}$  &$0.0562\pm0.0063$ \\
$\ln(10^{10} A_s)$ &$3.062\pm0.010$ &$3.05\pm0.011$ &$3.047\pm{0.011}$ &$3.039\pm0.017$ \\
$n_s$ &$0.9743\pm{0.0034}$ &$0.9724\pm{0.0037}$ &$0.9621\pm{0.0058}$ &$0.9640\pm{0.0067}$\\
$H_0 \, [\rm km \, s^{-1} \, Mpc^{-1}]$ &$68.30\pm0.27$ &$66.82^{+0.56}_{-0.54}$ &$69.42^{+0.52}_{-0.53}$ &$69.30^{+0.57}_{-0.55}$\\
$\Omega_K$ &$-$ &$-$ &$0.00059^{+0.00140}_{-0.00142}$ &$0.0006\pm0.0014$ \\
$A_{\rm lens}$ &$-$ &$-$ &$-$ &$1.025\pm0.037$ \\
$\Delta m_e /m_e$ &$-$ &$-$  &$0.0108^{+0.0057}_{-0.0056}$ &$0.0090\pm0.0062$ \\
$w_0$ &$-$ &$-0.771^{+0.055}_{-0.053}$ &$-$ &$-$ \\
$w_a$ &$-$ &$-0.72\pm0.21$ &$-$ &$-$ \\
\hline
$z_{*}$ &$1089.45\pm0.16$ &$1089.56\pm0.18$ &$1099.95\pm{5.36}$ &$1098.21\pm6.00$ \\
$r_{*}$ &$144.90\pm0.17$ &$144.72\pm0.22$ &$143.12\pm{0.76}$ &$143.43\pm0.87$ \\
$z_{\rm d}$ &$1060.16\pm0.23$ &$1060.15\pm0.23$ &$1069.55\pm{4.89}$ &$1068.01\pm5.46$ \\
$r_{\rm d}\, [\rm Mpc]$ &$147.51\pm0.19$ &$147.34\pm0.23$ &$145.77\pm{0.74}$ &$146.07\pm0.85$ \\
$\Omega_m$ &$0.3023\pm0.0036$ &$0.3175\pm0.0056$ &$0.3000\pm{0.0039}$ &$0.2996\pm0.0040$ \\
$S_8$ &$0.8133^{+0.0066}_{-0.0069}$ &$0.8253\pm0.0077$ &$0.8194\pm0.0074$ &$0.813\pm0.012$ \\
\hline
$\ln(Z)$ &$-1103.28\pm0.19$ &$-1099.99\pm0.21$ &$-1108.63\pm0.22$ &$-1110.98\pm0.22$ \\
$\ln \mathcal{B}$ &0 &$3.29\pm0.28$ &$-5.35\pm0.29$ &$-7.7\pm0.29$ \\
max$\ln(L)$ &-1068.80 &-1059.79 &-1065.61 &-1065.51 \\
$\Delta \rm AIC$ &0 &-14.01 &-2.36 &-0.60 \\
\hline
\end{tabular}
\caption{68\% credible intervals for the $\Lambda$CDM and several extended models, using Planck and ACT CMB power spectra, in combination with Planck PR4 and ACT DR2 CMB lensing reconstruction, BAO measurements from DESI DR2, CC measurements, and SNe Ia data sets from DES Y5. The top group (first 11 rows) lists the base parameters sampled with flat priors. The middle group lists derived parameters. The bottom group shows statistical metrics quantifying the fit efficiency.\label{tab:2}}
\end{sidewaystable*}

\begin{sidewaystable*}[htbp]
\renewcommand{\arraystretch}{1.3} 
\begin{tabular}{l|c@{\hspace{5.0em}}c@{\hspace{5.0em}}c@{\hspace{5.0em}}c}

\hline
\multicolumn{5}{c}{Planck 2018 + ACT DR6 + Lensing + DESI DR2 + CC + Pantheon+}\\
\hline
Parameter &$\Lambda \rm CDM$ &$w_0 w_a \rm CDM$ &$\Delta m_e + \Omega_K$  &$A_{\rm lens} + \Delta m_e + \Omega_K$\\
\hline
$\Omega_b h^2$ &$0.02264\pm0.00010$ &$0.02255\pm0.00011$ &$0.02258\pm{0.00011}$ &$0.02258\pm0.00012$\\
$\Omega_c h^2$ &$0.11673\pm0.00062$ &$0.11858\pm0.00087$ &$0.1226\pm{0.0015}$ &$0.12241\pm0.00174$\\
$\tau$ &$0.0639\pm0.0060$ &$0.0599\pm0.0059$  &$0.0555\pm{0.0057}$ &$0.0553\pm0.0061$\\
$\ln(10^{10} A_s)$ &$3.0671\pm0.0099$ &$3.052\pm0.011$ &$3.041\pm{0.011}$ &$3.038\pm0.017$\\
$n_s$ &$0.9765\pm{0.0034}$ &$0.9722\pm{0.0036}$ &$0.9554\pm{0.0056}$ &$0.9560\pm{0.0060}$\\
$H_0 \, [\rm km \,s^{-1} \, Mpc^{-1}]$  &$68.75\pm0.26$ &$68.98^{+0.53}_{-0.52}$ &$70.53^{+0.47}_{-0.46}$ &$70.51\pm0.49$\\
$\Omega_K$ &$-$ &$-$ &$-0.0001^{+0.0013}_{-0.0013}$ &$-0.0001\pm0.0013$\\
$A_{\rm lens}$ &$-$ &$-$ &$-$  &$1.008^{+0.038}_{-0.036}$\\
$\Delta m_e /m_e$ &$-$ &$-$ &$0.0203\pm0.0053$ &$0.020\pm0.0057$\\
$w_0$ &$-$ &$-0.894^{+0.052}_{-0.051}$ &$-$ &$-$\\
$w_a$ &$-$ &$-0.54^{+0.20}_{-0.21}$ &$-$ &$-$\\
\hline
$z_{*}$ &$1089.25\pm0.15$ &$1089.54\pm0.18$ &$1108.84\pm{5.01}$ &$1108.31\pm5.43$\\
$r_{\rm *}\, [\rm Mpc]$ &$145.09\pm0.17$ &$144.67\pm0.21$ &$141.98\pm{0.69}$ &$142.08\pm0.79$\\
$z_{\rm d}$ &$1060.27\pm0.22$ &$1060.22\pm0.23$ &$1077.83\pm{4.63}$ &$1077.39\pm4.97$\\
$r_{\rm d}\, [\rm Mpc]$ &$147.68\pm0.18$ &$147.28\pm0.22$ &$144.64\pm{0.69}$ &$144.74\pm0.78$\\
$\Omega_m$ &$0.2964\pm0.0034$ &$0.2981\pm0.0048$ &$0.2934\pm{0.0035}$ &$0.2932\pm0.0037$\\
$S_8$ &$0.8045^{+0.0066}_{-0.0067}$ &$0.8186^{+0.0077}_{-0.0076}$ &$0.8151^{+0.0073}_{-0.0072}$ &$0.813\pm0.012$\\
\hline
$\ln(Z)$ &$-1020.15\pm0.19$ &$-1021.81\pm0.21$ &$-1018.49\pm0.22$ &$-1020.83\pm0.22$\\
$\ln \mathcal{B}$ &0 &$-1.66\pm0.28$ &$1.66\pm0.29$  &$-0.68\pm0.29$ \\
max$\ln(L)$ &-985.86 &-981.75 &-975.15 &-975.51 \\
$\Delta \rm AIC$ &0 &-4.23 &-17.42 &-14.71 \\
\hline
\end{tabular}
\caption{Parameter 68\% intervals for the $\Lambda$CDM and several extended models from Planck and ACT CMB power spectra, in combination with Planck PR4, ACT DR2 CMB lensing reconstruction, BAO measurements from DESI DR2, CC measurements, and SNe Ia data sets from Pantheon+. The top group of 11 rows are the base parameters, which are sampled in the Bayesian analysis with flat priors. The middle group lists derived parameters. The bottom group shows statistical metrics quantifying the fit efficiency.\label{tab:3}}
\end{sidewaystable*}

\begin{sidewaystable*}[htbp]
\renewcommand{\arraystretch}{1.3} 
\begin{tabular}{l|c@{\hspace{5.0em}}c@{\hspace{5.0em}}c@{\hspace{5.0em}}c}

\hline
\multicolumn{5}{c}{Planck 2018 + ACT DR6 + Lensing}\\
\hline
Parameter &$\Lambda \rm CDM$ &$w_0 w_a \rm CDM$ &$A_{\rm lens} + \Omega_K$  &$A_{\rm lens} + \Delta m_e + \Omega_K$\\
\hline
$\Omega_b h^2$ &$0.02250\pm0.00011$ &$0.02253\pm0.00011$ &$0.02255\pm0.00012$ &$0.02258\pm0.00012$\\
$\Omega_c h^2$ &$0.1190\pm0.0011$ &$0.1185\pm0.0012$ &$0.1180\pm0.0014$ &$0.1202\pm0.0018$\\
$\tau$ &$0.0600\pm0.0062$ &$0.0585\pm0.0060$ &$0.0577\pm0.0063$ &$0.0561\pm0.0060$\\
$\ln(10^{10} A_s)$ &$3.055\pm0.011$ &$3.043\pm0.012$ &$3.038\pm0.018$ &$3.035\pm0.017$\\
$n_s$ &$0.9714\pm0.0040$ &$0.9722\pm0.0040$ &$0.9732\pm0.0044$ &$0.9640\pm0.0067$\\
$H_0 \, [\rm km \,s^{-1} \, Mpc^{-1}]$ &$67.77^{+0.48}_{-0.47}$ &$85.58^{+10.01}_{-12.94}$ &$67.22^{+9.13}_{-8.68}$ &$65.91^{+8.93}_{-8.64}$\\
$\Omega_K$ &$-$ &$-$ &$-0.0019^{+0.0155}_{-0.0236}$ &$-0.0083^{+0.0173}_{-0.0264}$\\
$A_{\rm lens}$ &$-$ &$-$ &$1.031^{+0.13}_{-0.12}$ &$0.98^{+0.13}_{-0.12}$\\
$\Delta m_e /m_e$ &$-$ &$-$ &$-$ &$0.0116^{+0.0066}_{-0.0063}$\\
$w_0$ &$-$ &$-1.30^{+0.55}_{-0.47}$ &$-$ &$-$ \\
$w_a$ &$-$ &$-1.10^{+1.67}_{-1.34}$ &$-$ &$-$ \\
\hline
$z_{*}$ &$1089.63\pm0.21$ &$1089.56\pm0.21$ &$1089.49\pm0.24$ &$1100.52\pm6.13$\\
$r_{\rm *}\, [\rm Mpc]$ &$144.61\pm0.27$ &$144.71\pm0.27$ &$144.81\pm0.32$ &$143.31\pm0.88$\\
$z_{\rm d}$ &$1060.14\pm0.23$ &$1060.16\pm0.23$ &$1060.18\pm0.24$ &$1070.30\pm5.65$\\
$r_{\rm d}\, [\rm Mpc]$ &$147.23\pm0.27$ &$147.32\pm0.28$ &$147.42\pm0.32$ &$145.95\pm0.86$\\
$\Omega_m$ &$0.3096\pm0.0066$ &$0.194\pm0.060$ &$0.313\pm0.086$ &$0.330\pm0.093$\\
$S_8$ &$0.824^{+0.011}_{-0.010}$ &$0.769^{+0.040}_{-0.029}$ &$0.819^{+0.104}_{-0.092}$ &$0.841^{+0.093}_{-0.082}$\\
\hline
$\ln(Z)$ &$-264.35\pm0.19$ &$-263.75\pm0.19$ &$-268.41\pm0.20$ &$-270.94\pm0.22$\\
$\ln \mathcal{B}$ &0 &$0.6\pm0.27$ &$-4.06\pm0.28$ &$-6.59\pm0.29$ \\
max$\ln(L)$ &-230.84 &-229.93 &-230.23 &-228.76\\
$\Delta \rm AIC$ &0 &2.18 &2.78 &1.82 \\
\hline
\end{tabular}
\caption{Parameter 68\% intervals for the $\Lambda$CDM and several extended models from Planck and ACT CMB power spectra, in combination with Planck PR4 and ACT DR2 CMB lensing
reconstruction. The top group of 11 rows are the base parameters, which are sampled in the Bayesian analysis with flat priors. The middle group lists derived parameters. The bottom group shows statistical metrics quantifying the fit efficiency.\label{tab:4}}
\end{sidewaystable*}

\begin{sidewaystable*}[htbp]
\renewcommand{\arraystretch}{1.3} 
\begin{tabular}{l|c@{\hspace{5.0em}}c@{\hspace{5.0em}}c@{\hspace{5.0em}}c}
\hline
\multicolumn{5}{c}{Planck HiLLiPoP + Lensing + DESI DR2}\\
\hline
Parameter &$\Lambda \rm CDM$ &$w_0 w_a \rm CDM$  &$A_{\rm lens} + \Omega_K$ &$A_{\rm lens} + \Delta m_e + \Omega_K$\\
\hline
$\Omega_b h^2$ &$0.02237\pm{0.00011}$ &$0.02226\pm{0.00011}$ &$0.02234\pm{0.00013}$ &$0.02230\pm{0.00013}$\\
$\Omega_c h^2$  &$0.11725\pm{0.00061}$ &$0.11887\pm{0.00088}$ &$0.1179\pm{0.0011}$ &$0.1163\pm{0.0018}$\\
$\tau$ &$0.0612\pm{0.0060}$ &$0.0581\pm{0.0059}$ &$0.0574\pm{0.0060}$ &$0.0583\pm{0.0061}$\\
$\ln(10^{10} A_s)$ &$3.061\pm{0.010}$ &$3.048\pm{0.011}$ &$3.038\pm{0.017}$ &$3.041\pm{0.017}$\\
$n_s$ &$0.9718\pm{0.0031}$ &$0.9679\pm{0.0034}$ &$0.9704\pm{0.0040}$  &$0.9763\pm{0.0065}$\\
$H_0 \, [\rm km \,s^{-1} \, Mpc^{-1}]$ &$68.36\pm0.27$ &$64.25^{+1.89}_{-1.77}$ &$68.65\pm0.31$ &$67.94\pm0.67$\\
$\Omega_K$ &$-$ &$-$ &$0.0012\pm0.0011$ &$0.0025^{+0.0016}_{-0.0015}$\\
$A_{\rm lens}$ &$-$ &$-$ &$1.056^{+0.040}_{-0.037}$ &$1.075^{+0.044}_{-0.043}$\\
$\Delta m_e /m_e$ &$-$ &$-$ &$-$ &$-0.0095^{+0.0078}_{-0.0079}$\\
$w_0$ &$-$ &$-0.52^{+0.21}_{-0.20}$ &$-$ &$-$\\
$w_a$ &$-$ &$-1.37^{+0.57}_{-0.61}$ &$-$ &$-$\\
\hline
$z_{*}$ &$1089.64\pm{0.16}$ &$1089.94\pm{0.21}$ &$1089.75\pm{0.25}$ &$1080.77\pm{7.54}$\\
$r_{\rm *}\, [\rm Mpc]$ &$145.16\pm{0.16}$ &$144.82\pm{0.20}$ &$145.01\pm{0.25}$ &$146.24\pm{1.07}$\\
$z_{\rm d}$ &$1059.72\pm{0.23}$ &$1059.59\pm{0.23}$ &$1059.70\pm{0.24}$ &$1051.40\pm{6.90}$\\
$r_{\rm d}\, [\rm Mpc]$ &$147.83\pm{0.17}$ &$147.53\pm{0.21}$ &$147.69\pm{0.25}$ &$148.90\pm{1.05}$\\
$\Omega_m$ &$0.3004\pm{0.0035}$ &$0.343\pm{0.020}$ &$0.2992\pm{0.0037}$ &$0.3020\pm{0.0045}$\\
$S_8$ &$0.8088\pm0.0067$ &$0.838^{+0.012}_{-0.012}$ &$0.801^{+0.012}_{-0.011}$ &$0.799\pm0.012$\\
\hline
$\ln(Z)$ &$-2862.67\pm0.27$ &$-2863.69\pm0.28$ &$-2868.43\pm0.29$ &$-2871.62\pm0.30$\\
$\ln{\mathcal{B}}$ &0 &$-1.02\pm0.39$ &$-5.76\pm0.40$ &$-8.95\pm0.40$\\
max$\ln(L)$ &-2792.53 &-2790.40 &-2791.85 &-2792.23\\
$\Delta \rm AIC$ &0 &-0.26 &2.64 &5.40 \\
\hline
\end{tabular}
\caption{Parameter 68\% intervals for the $\Lambda$CDM and several extend models from Planck PR4 (HiLLiPoP), CMB lensing and DESI DR2 reconstruction. The top group of 11 rows are the base parameters, which are sampled in the Bayesian analysis with flat priors. The middle group lists derived parameters. The bottom group shows the statistic factors that analysing the fitting efficiency quantitatively.\label{tab:5}}
\end{sidewaystable*}

\begin{sidewaystable*}[htbp]
\renewcommand{\arraystretch}{1.3} 
\begin{tabular}{l|c@{\hspace{5.0em}}c@{\hspace{5.0em}}c@{\hspace{5.0em}}c}
\hline
\multicolumn{5}{c}{Planck NPIPE + Lensing + DESI DR2}\\
\hline
Parameter &$\Lambda \rm CDM$ &$w_0 w_a \rm CDM$  &$A_{\rm lens} + \Omega_K$ &$A_{\rm lens} + \Delta m_e + \Omega_K$\\
\hline
$\Omega_b h^2$ &$0.02235\pm{0.00012}$ &$0.02223\pm{0.00013}$ &$0.02230\pm{0.00014}$ &$0.02227\pm{0.00016}$\\
$\Omega_c h^2$ &$0.11740\pm{0.00062}$ &$0.11912\pm{0.00087}$ &$0.1185\pm{0.0012}$ &$0.1179\pm{0.0020}$\\
$\tau$ &$0.0608\pm{0.0059}$ &$0.0569\pm{0.0060}$ &$0.0565\pm{0.0060}$ &$0.0569\pm{0.0063}$\\
$n_s$ &$0.9690\pm{0.0033}$ &$0.96490\pm{0.00368}$ &$0.9667\pm{0.0042}$ &$0.9687\pm{0.0067}$\\
$H_0 \, [\rm km \,s^{-1} \, Mpc^{-1}]$ &$68.27^{+0.28}_{-0.27}$ &$64.08^{+2.02}_{-1.90}$ &$68.62^{+0.33}_{-0.32}$ &$68.34^{+0.80}_{-0.82}$\\
$\Omega_K$ &$-$ &$-$ &$0.0017^{+0.0011}_{-0.0012}$ &$0.0022^{+0.0019}_{-0.0017}$\\
$A_{\rm lens}$ &$-$ &$-$ &$1.048^{+0.037}_{-0.036}$ &$1.053^{+0.042}_{-0.040}$\\
$\Delta m_e /m_e$ &$-$ &$-$ &$-$ &$-0.0036^{+0.0095}_{-0.0099}$\\
$w_0$ &$-$ &$-0.49^{+0.22}_{-0.22}$ &$-$ &$-$\\
$w_a$ &$-$ &$-1.49^{+0.59}_{-0.65}$ &$-$ &$-$\\
\hline
$z_{*}$  &$1089.69\pm{0.18}$ &$1090.00\pm{0.21}$  &$1089.85\pm{0.26}$ &$1086.47\pm{9.14}$\\
$r_{\rm *}\, [\rm Mpc]$ &$145.14\pm{0.16}$ &$144.77\pm{0.20}$ &$144.90\pm{0.26}$ &$145.35\pm{1.29}$\\
$z_{\rm d}$ &$1059.67\pm{0.26}$ &$1059.55\pm{0.27}$ &$1059.65\pm{0.28}$ &$1056.45\pm{8.55}$\\
$r_{\rm d}\, [\rm Mpc]$ &$147.82\pm{0.18}$ &$147.48\pm{0.21}$ &$147.59\pm{0.26}$ &$148.04\pm{1.28}$\\
$\Omega_m$ &$0.3014\pm{0.0036}$ &$0.346\pm{0.022}$ &$0.3005\pm{0.0038}$ &$0.3017\pm{0.0050}$\\
$S_8$ &$0.8101^{+0.0068}_{-0.0063}$  &$0.840^{+0.012}_{-0.012}$ &$0.805\pm0.012$ &$0.80^{+0.013}_{-0.012}$\\
\hline
$\ln(Z)$ &$-5367.31\pm0.22$ &$-5367.45\pm0.23$ &$-5372.19\pm0.24$ &$-5375.31\pm0.25$\\
$\ln{\mathcal{B}}$ &0 &$-0.14\pm0.32$ &$-4.88\pm0.33$ &$-8.00\pm0.38$\\
max$\ln(L)$ &-5322.23 &-5318.16 &-5319.19 &-5319.83\\
$\Delta \rm AIC$ &0 &-4.14 &-2.08 &1.20\\
\hline
\end{tabular}
\caption{Parameter 68\% intervals for the $\Lambda$CDM and several extend models from Planck PR4 (NPIPE), CMB lensing and DESI DR2 reconstruction. The top group of 11 rows are the base parameters, which are sampled in the Bayesian analysis with flat priors. The middle group lists derived parameters. The bottom group shows the statistic factors that analysing the fitting efficiency quantitatively.\label{tab:6}}
\end{sidewaystable*}

\acknowledgments

We thank the anonymous referee for helpful comments and
suggestions. This work is supported in part by NSFC under grants of No. 12233011 and 12303056, and the Postdoctoral Fellowship Program of CPSF (No. GZB20250738).

\bibliography{biblio}
\bibliographystyle{JHEP}

\end{document}